# Liquid interface shaping and transport phenomena induced by spatially inhomogeneous vibrations

Benjamin Apffel, Christian Wilkinson and Emmanuel Fort

*Institut Langevin, ESPCI Paris, Université PSL, CNRS, 75005 Paris, France*

Vibrations can dynamically stabilize otherwise unstable liquid interfaces and produce new dynamic equilibria, called vibro-equilibria. Typically, the vibrations are homogeneous in the liquid and the liquid interface remains approximately flat. Here, we produce controlled vertical vibration gradients by taking advantage of the resonant oscillations of sinking submerged bubbles. The locally increased amplitude of the vibrations induces a local elevation of the liquid interface that can be controlled and engineered. The mean elevation of the interface can be linked theoretically with the local vibration amplitude by a simple formula that is tested experimentally. In addition, the transport of a floating body at the interface can be induced by secondary flows triggered by the amplitude gradients of the liquid vibrations.

## Introduction

Under classical experimental conditions, large liquid-air surfaces are horizontal as gravity tends to suppress any height difference at the interface. By applying additional forcing, it is however possible to induce deformations. A convenient way to apply such forcing is to accelerate the fluid, giving rise to inertial forces. Apart from rotation that gives the interface a parabolic shape [1], high frequency mechanical vibrations are of particular interest for interface deformation. The first observation of this phenomenon was reported by Faraday who noticed that droplets under vibration were flattened when compared to their shape at rest [2]. Decades later, Stephenson [3] and Kapitza [4] showed that a solid pendulum shaken hard enough at a frequency much larger than its natural frequency could be stabilized dynamically in its upright position. Since then, this idea has been extended to fluids in order to create new shapes of interfaces named vibro-equilibria. Those unexpected positions include stabilized horizontal interfaces with a lighter liquid under a denser one when the forcing is vertical as shown by G. H. Wolf in early experiments [5, 6]. In the same study, it was also shown that non-horizontal interfaces or stationnary wave-like patterns (called frozen waves) can be observed when the forcing is horizontal. Since then, many experimental, theoretical and numerical studies were performed in order to quantitatively understand the influence of vibrations on the shape of the interfaces. [7–15]

From an experimentalist perspective, some unwanted effects can hinder the appearance of the desired vibro-equilibrium. With water for instance, applying vibrations at a few tens of hertz with an amplitude similar to $g \sim 10$ ms$^{-2}$ leads to Faraday instability [2, 16–18] and even to droplet ejection from the interface, while this amplitude is too small to obtain significant interface deformation. To overcome this difficulty, many experiments were performed in micro- gravity environment in order to increase the relative ratio of vibrational effects compared to gravity [9, 12, 13]. The use of viscous fluids makes it possible to delay the instability thresholds and thus to increase the vibration amplitudes. For instance, a viscosity a thousand times higher than that of water, obtained with silicone oil for example, makes it possible to apply important forcing, typically tens of g at 100 Hz, without destabilizing the interface. Under these conditions, non-horizontal vibro-equilibrium positions in arbitrary directions have recently been observed [19].

In all the mentioned experiments, the forcing of the fluid is spatially uniform as the whole fluid is shaken with the same acceleration. Here, we propose to shape the liquid interface using a controlled inhomogeneous vibration field. We take advantage of the atypical behavior of air bubbles in a vibrated tank that can sink providing the forcing is strong enough [20, 21]. When trapped at the bottom of the container, the bubbles act as mechanical springs and can locally increase the forcing amplitude. This position-dependent forcing induces a local deformation of the interface.

This article is organized as follows. We first describe the experimental setup and the effect of sinking bubbles to induce amplitude gradients of the vibrations at the liquid interface. We then propose a model to relate the liquid interface relative height to the amplitude of the local shaking and we compare the model to experimental results. The next section shows how it is possible to shape the interface by controlling the positions of the bubbles. The last section is devoted to transport phenomena driven by secondary flows in the liquid layer for floaters placed on the interface.

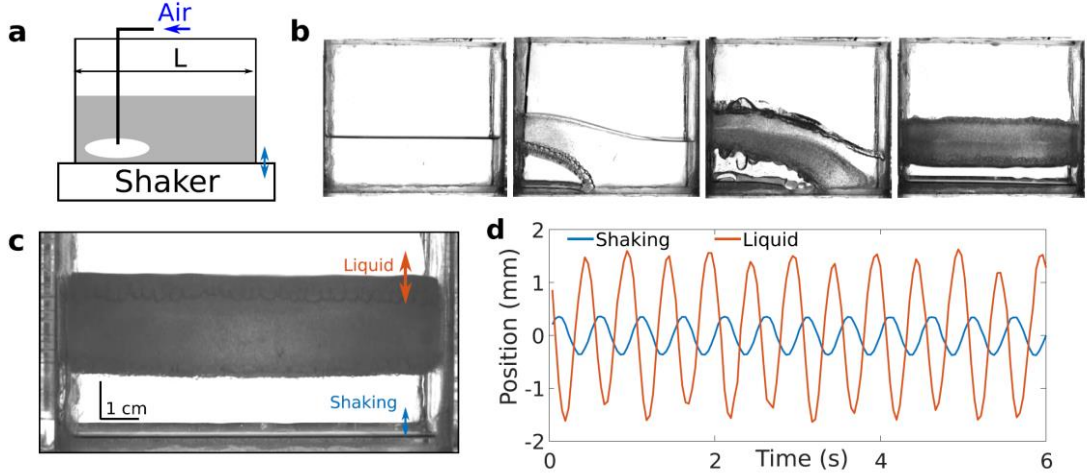

**Fig. 1.** (a) Experimental setup: a bath of length $L = 10$ cm and width 4 cm filled with silicon oil, vibrated at a frequency of $f_{exc} = 100$ Hz with a shaker. Sinking air bubbles are injected through a syringe under strong shaking. (b) Snapshots of the creation of an air layer at the bottom of a vibrated tank by inflating a small air bubble. Starting from a small air bubble, the injection of more air eventually leads to a complete air layer trapped under the liquid layer. The interface is stabilized with the vertical vibrations (c) Snapshot of a movie from which the vertical position of the liquid and the container is extracted. (d) Strobbed vertical positions as a function of time of the container and the liquid layer obtained from a movie taken at 25.5 frame/second.

## 1. Experimental stabilization of unusual liquid-air interfaces
## 1.1. From sinking air bubbles to interface stabilization and resonance

Rayleigh-Taylor instability occurs when a denser liquid is placed above a less dense one [22–24]. This occurs typically when paint is spread on a ceiling or when a glass of liquid is turned upside down, resulting in the fall of the liquid. It has been shown that high frequency vibrations can be used to stabilize such instability [5, 6]. The underlying theoretical description can be summarized as follow. A perfectly horizontal liquid-air interface (with liquid above) is an equilibrium position as the interface is orthogonal the hydrostatic pressure gradient. However, this equilibrium is unstable as any deformation larger than the capillary length tends to grow exponentially with time. The unstable behavior of a sinusoidal interface perturbation of wave vector $k$ is associated in the linear regime to an equation of the form $\ddot{X}_k - \omega_0^2(k)X_k = 0$ with $X_k$ the deformation associated to wave vector $k$ and $\omega_0^2 = gk - \gamma k^3/\rho > 0$ for $k$ small enough. This equation is also the evolution equation of a pendulum around its unstable equilibrium position, $X$ being the angular deflection of the pendulum with respect to this equilibrium position and $\omega 0$ its pulsation. It was shown by Kapitza [4] that such unstable pendulum position can be stabilized by modulating gravity along time through shaking. As a consequence, such liquid air interface can be stabilized if all the unstable modes occurring on the bath are stabilized by the vibration.

Such an inverted configuration can be achieved in different ways. A first possibility is to start with the liquid in the bottom of a tank, vibrate the container and rotate the whole experiment in order to get the liquid above the air. Alternatively, we propose to use the atypical behaviour of air bubbles in vibrated liquids [25]. Due to its compressibility, an air bubble in such media will experience a force arising from vibrations which can overcome its buoyancy [20]. As a consequence, if one injects air in the liquid through a syringe as shown in figure 1a, it will sink toward the bottom of the tank if the shaking is strong enough. If one now keeps inflating the bubble, it will grow without destabilizing the liquid layer as Rayleigh-Taylor instability is suppressed by the vibration. At the end, the air will eventually fill the container bottom entirely [21] as shown in figure 1b. In this case, the liquid-air interface is horizontal and stable as expected from the previous considerations. Figure 1c shows a 10 cm wide liquid layer stabilized via the vibrations above an air layer that has been injected. The liquid is silicon oil with kinematic viscosity $\nu = 1000$ cSt to avoid

Faraday instability and the bath is vibrated at $f_{exc}$ = 100 Hz. An interesting side effect of the vibration is the absorption of an increased number of air bubbles in the liquid compared to what occurs at rest [25]. When placed in front of a LED panel, the liquid thus appears darker than the illuminated background as shown 1b where the amount of bubbles progressively increases with time. When completely saturated, the liquid appears opaque as in figure 1c which provides an efficient method to measure its position.

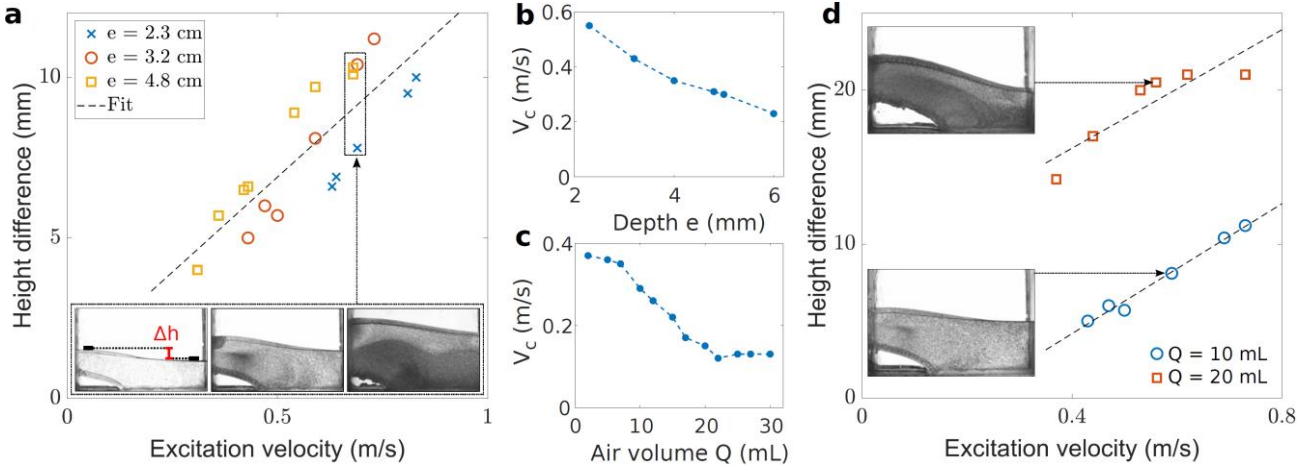

**Fig. 2.** (a) Height difference between the two extremal parts of the interface (see inset) for different excitation velocity and different initial depths $e$ = 2.3 cm (cross), $e$ = 3.2 cm (circles), $e$ = 4.8 cm (squares) of liquid. The volume of air $Q$ = 10 ml is constant for all measurements. The dashed line is a guide for the eye. Inset: pictures of the interface for fixed excitation velocity $A\omega \approx 0.6$ ms$^{-1}$ and different depth $e$. (b) Minimal excitation velocity $V_C$ for an air volume of Q=10 ml and different liquid depth $e$ (blue points are experiments; dashed line is a guide for the eye). (c) Measured $V_C$ for different air volume $Q$ and constant depth $e$ = 5 cm. (d) Height difference for different excitation velocity when the initial depth $e$ = 3.2 cm is constant. The volume of air is varied from $Q$ = 10 ml (circles) to $Q$ = 20 ml (squares). Dashed lines are guides for the eye. Insets: pictures for excitation velocity $A\omega \approx 0.9$ ms$^{-1}$ and different volumes of air.

Interestingly, the air layer trapped between the bottom and the liquid layer acts as a spring-mass oscillator with a resonance frequency close to the excitation frequency [26]. Consequently, the vibration amplitude of the liquid is larger and generally not in phase with the mechanical excitation. The relative phase between the input and the output increases with the excitation frequency and goes from $\Phi$ = 0 to $\Phi$ = $\pi$ when the excitation frequency crosses the resonant frequency, the phase being $\Phi$ = $\pi/2$ at resonance. To characterize this motion, a movie at 25.5 fps is taken while the shaking is kept at constant amplitude and frequency $f_{exc}$ = 100 Hz. As the liquid appears darker than the background, usual detection algorithms can be applied on each image to find the liquid layer's barycentric vertical position as well as the container's position. The strobed oscillation as a function of time of the vertical positions of the liquid layer and the container (with the mean value subtracted) are shown in figure 1d. Both curves present an almost perfect sinusoidal oscillation, the slow apparent evolution of the oscillation being due to the choice of the sampling frequency of 25.5 fps compared to the excitation frequency of 100 Hz. This allows to precisely estimate the oscillation amplitude of the container $A$ = 0.4 mm and the oscillation amplitude of the liquid $A_l$ = 1.5 mm. The relative phase between the liquid oscillation and the container one is $\Phi \approx 3.6$ rad. As mentioned before, this indicates that the shaking frequency is larger than the resonance frequency of the equivalent spring-mass oscillator. As $\Phi$ is slightly larger than $\pi$, the harmonic oscillator model that predicts a phase difference of $0 < \Phi < \pi$ does not completely describes the dynamic of the system in these experimental conditions.

When looking at figure 1b, it can be seen that when only part of the bottom is filled with air, non-horizontal interfaces can appear. We now propose to investigate this type of deformations for various experimental parameters.

## 1.2. Bending interfaces with partial air layers

In the previous section, the forcing applied to the fluid was amplified by the air layer but uniform over the whole interface. We now focus on the case where the fluid experiences non-uniform forcing. Experimentally, this occurs when the air layer only fills a significant fraction of the bottom of the container. In this case, the liquid interface is not horizontal but presents a bended shape above the partial layer of air as shown in figure 2. For a fixed injected volume Q = 10 ml of air, the height difference at the interface is measured for different forcing velocity $A\omega$. The results in figure 2a show that the height difference increases with the forcing velocity. The applied velocities were restricted to $A\omega < 0.8$ m/s as greater forcing induced Faraday instability above the air layer.

Performing the same experiment for different depth of liquid $e$ shows that the height difference only marginally depends on the amount of liquid characterized by the depth $e$. This depth is however a crucial parameter for the stability of the layer. Bellow a critical velocity $V_c$, the air layer was observed going upward again. The dependency of $V_c$ with respect to $e$ is plotted in figure 2b and shows that it decreases when the amount of liquid increases. Given a fixed depth $e = 5$ cm, we also measured the critical velocity $V_c$ for different air volumes Q as shown in figure 2c. The critical velocity $V_c$ goes from constant value $V_c \approx 0.35$ m/s for small volumes Q < 5 ml to another smaller constant value $V_c \approx 0.12$ m/s for large air volumes $Q > 20$ ml. A quantitative interpretation of those results would require taking into account the resonance of the bubble and its interaction with the vibration amplitude and the surrounding liquid. Such study is beyond the scope of the present article. Our measurements nevertheless give experimental insight on the forcing strength dependency with respect to different parameters.

When the volume of air Q is varied from 10 ml to 20 ml ($e = 3.2$ cm being fixed), we also observe the height difference to increases significantly as shown in figure 2b. For the larger forcing, the height difference reaches 2 cm, to be compared with the liquid thickness $e = 3.2$ cm. In both cases, the height difference increases linearly within the excitation velocity range applied.

The stabilization of these interfaces is of different type than the previous Rayleigh-Taylor stabilization. In fact a bended interface does not seems correspond to an equilibrium position as the interface is not orthogonal to the hydrostatic pressure. We show in the next section that the spatially inhomogeneous forcing is directly linked to these non-horizontal equilibrium interfaces.

## 2. Connection between the height difference and non-uniform motion

We now aim to investigate the link between non-homogeneous spatial excitation and the resulting interface deformation. By measuring the vertical displacement along time of different points of the interface as in figure 3a-b, we see that the interface above the air layer oscillates with greater amplitude than the interface part with no air bellow. Far from the air layer the interface oscillation is exactly the mechanical oscillation imposed by the shaker. Contrary to section 2.1, the liquid above the air oscillates in phase with the excitation. Two factors can explain this difference. As the air layer dimensions are smaller, the resonance frequency of the spring-mass oscillator is expected to be larger what tends to decrease the phase difference $\Phi$. Moreover, the high viscosity of the silicon oil tends to prevent the apparition of phase difference that would cause large shears in the fluid. The non-uniform air layer thus provides spatially non-homogeneous but in phase forcing as well as non-horizontal interface. The same kind of observations but with phase differences were observed in the case of non-horizontal forcing [19].

Those two effects are related and can be analytically connected. We assume that the interface has equation $y = h(x, t)$. If one neglect capillarity, the pressure boundary condition at the interface imposes

$$P(x, h(x, t)) = P_{\text{atm}} \tag{1}$$

where $P_{\text{atm}}$ is the atmospheric pressure. Due to shaking, all quantities including the pressure, the interface height $h(x, t)$ and the fluid velocity in the laboratory frame $v(x, t)$ at the interface will oscillate at the excitation frequency $f_{\text{exc}}$. We also define the fast oscillating velocity amplitude $V(x)$ so that $v(x, t) = V(x) \sin(2\pi f_{\text{exc}} t)$. In the same manner as in previous section, we can define mean variables over one fast excitation period such as the mean interface position $\bar{h}(x)$. Taking the mean of equation (1) gives [10, 11]

$$\tfrac{1}{4} V^2(x) - g\bar{h}(x) = C \tag{2}$$

where $C$ is a constant. This equation provides a simple connection between the fast oscillating fluid velocity amplitude $V(x)$ and the mean height of the interface $\bar{h}(x)$. If we take two points at the interface with fast velocity $V_1, V_2$ and mean height $\bar{h}(x)$ and $\bar{h}(x)$, we get the difference $\Delta \bar{h} = \bar{h}_2 - \bar{h}_1$ as

$$\Delta \bar{h} \doteq \bar{h}_2 - \bar{h}_1 = \frac{1}{4g}(V_2 - V_1) \tag{3}$$

The condition (2) ensures that the interface is a constant pressure area and so at equilibrium. As a consequence, a non-horizontal interface can only be an equilibrium position if the velocity along the interface is non-uniform. This is in strong contrast with section 2.1 where the isobar lines where always horizontal. Moreover, this condition does not imply the stability of the equilibrium nor describes the equilibrium shape but only predicts the height difference between two areas with different velocities. The stability of the equilibrium can nevertheless be assumed to be realized through vibrations as for Rayleigh-Taylor stabilization. In certain cases, the exact shape of the interface can be computed numerically by minimizing an averaged Lagrangian [13,14].

We now propose to verify condition (2) experimentally. When the interface forms small angles with respect to the horizontal, one can deduce $V(x)$ from the interface position. For each point $x$ of the interface corresponding to a fixed column of the image 3a, we detect the vertical displacement along time (see inset in figure 3b). From this, we deduce the fast oscillation amplitude $A(x)$. As the measurements are quite noisy, a moving mean filter of total width 5 mm is applied on the original $V(x)$ to smooth the result with respect to $x$. From equation (2) we can predict a height profile $\bar{h}_v(x)$ associated to the measured velocity. This profile is compared in figure 3b to the mean profile $\bar{h}(x)$ directly measured on the images. Both present reasonable agreement and show that the introduction of the air layer induces inhomogeneous spatial forcing resulting in non-horizontal interface. Even though the data have been smoothed, the prediction resulting from the velocity measurements exhibits oscillations and irregularities that prevent a complete quantitative comparison.

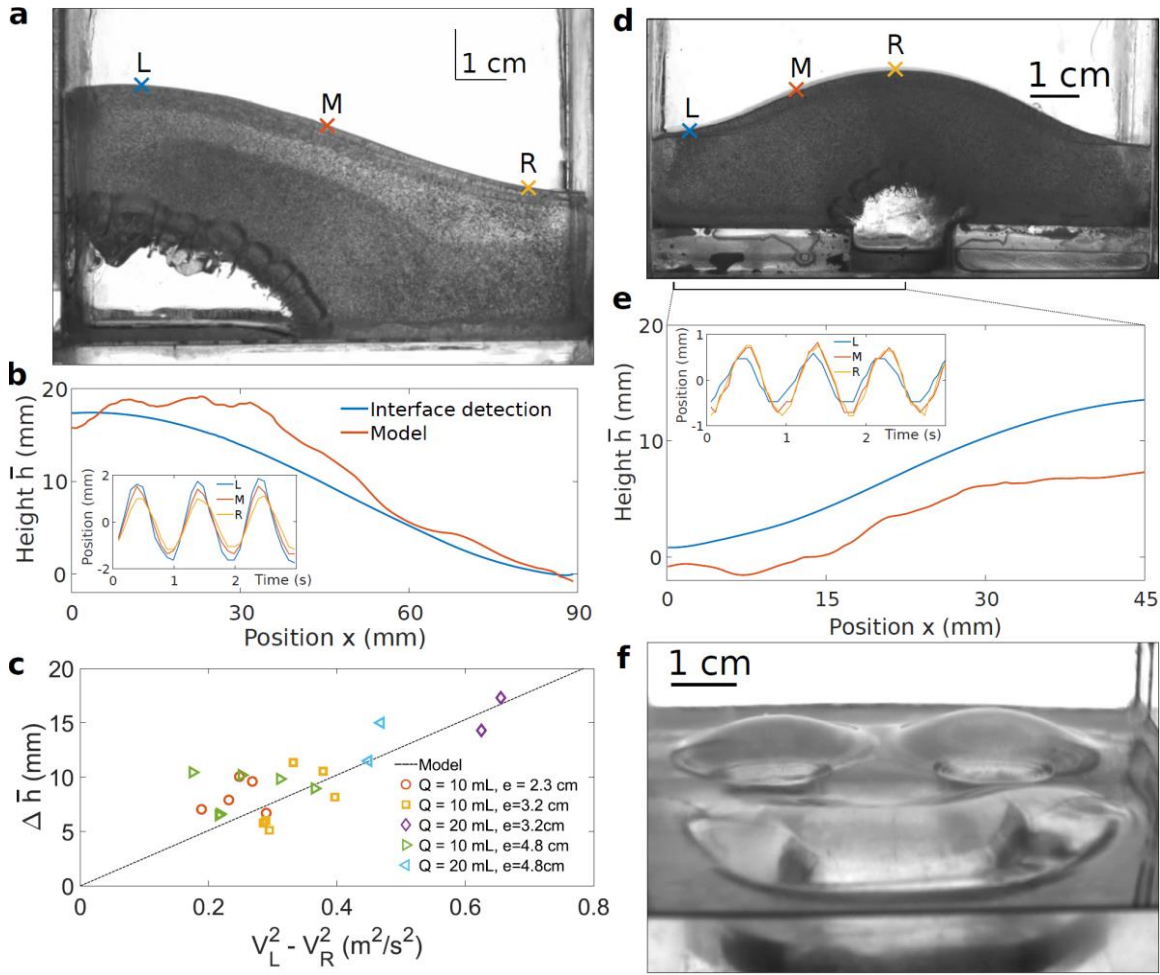

**Fig. 3.** (a) Picture of a liquid interface bended by a volume of air Q = 20 ml trapped in the bottom. The initial depth of liquid is $e = 3.2$ cm and the excitation amplitude is $A\omega = 0.7$ ms$^{-1}$. (b) Mean interface profile detected on a movie (blue line) and prediction (red line) from equation (2). The velocity $V(x)$ is obtained by measuring the vertical displacement of each point of the interface. Inset: strobed oscillation of points L, M, R along time at 10.1 frame/second. (c) Height difference $\Delta h$ as a function of velocity difference $V_L^2 - V_R^2$ at both sides of the tank for various experimental conditions and associated prediction from equation (3). (d) Interface deformation by an air layer trapped in a bottom defect in the center of the container. Frequency shaking is 178 Hz. (e) Measured interface (blue line) and prediction from equation (2) using the measured velocity profile (red line). Inset: strobbed position of different points of the interface along time. (f) Interface deformation induced by air trapped in a smiley-shaped bottom with shaking frequency of 250 Hz

More generally, we can test the model on the data presented in figure 2. For simplicity, we only focus on the difference between the left and right side above and far from the air layer. The equation (3) provides a linear relation between the fast velocity inhomogeneity $V_L^2 - V_R^2$ and the elevation difference $\Delta \bar{h}$ with slope $1/4g$. The quantities $V$ and $h$ measured at each side are mean values taken on a width of 6 mm. The results are presented in figure 3c as well as theoretical prediction. The measurement presents some dispersion due to measurement uncertainties but exhibits the trend predicted by the model.

## 3. Control of the air layer position

For now, the air layer position was not chosen but imposed by the system. When air is injected, is systematically tends to drift toward the closest edge (the left wall in our case). As a consequence, the interface could only be deformed near a wall. Introducing defects in the bottom profile makes possible to trap air at arbitrary positions. Figure 3d shows a typical example of an air layer trapped in the hole made in the center of the tank. As the size of the layer is smaller, the shaking frequency was increased to $f = 178$ Hz to obtain significant interface deformation. We performed the same interface analysis as before to extract the interface velocity and deduce the interface profile predicted by (3). We still observe larger velocities where the interface is higher but the model underestimates the height difference. Looking at the temporal position of different points along time (inset in figure 3e) shows that the signals are less sinusoidal than in figure 3b leading to less precise velocity measurement. This suggests that either the interface detection failed to properly measure the interface displacement in this case or that non-linearity starts to occur for these range of parameters. As a consequence, equation (3) fails to predict correctly the height difference.

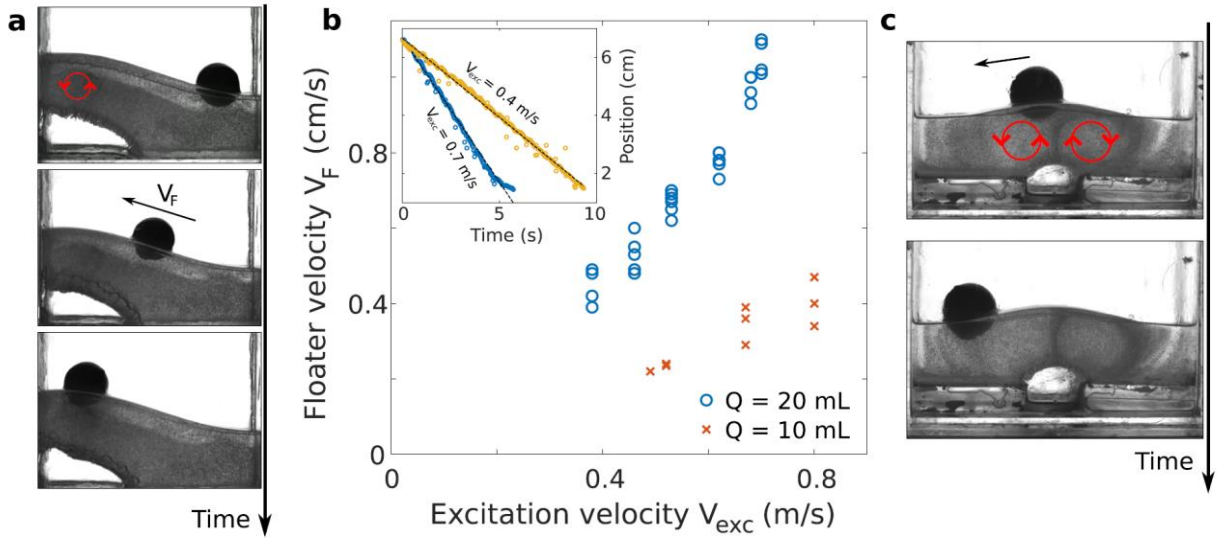

**Fig. 4.** (a) Pictures of a spherical floater with mass $m = 0.7$ g. When placed near the air layer, the floater gets attracted toward the highest part of the interface. A vortex (red circle) is generating a surface flow in the same direction. (b) Floater velocity during the ascent for different excitation velocity and two volumes of air layer Q = 10 ml (red crosses) and Q = 20 ml (blue circle). Each point corresponds to one experimental realization. Inset: horizontal position of the floater along time. The velocity $V_F$ is extracted from a linear fit. (c) Picture of the same floater when placed above an air layer trapped in the center. The induced vertices push the floater down to the lower part of the interface.

We can nevertheless use this idea to implement arbitrary interface deformations by properly engineering the bottom profile. Figure 3f shows the case of a smiley-shaped bottom with each cavity filled with air vibrated at $f = 250$ Hz. Different air layers do not seem to interact unless some air is injected in an already full of air defect. When the air starts to escape, it will leak toward the surface, one of the edges of the container or interact with the other air layers depending on the forcing and the configuration.

We have shown here that the interface deformation is closely related to spatially inhomogeneous forcing. Shaping properly the bottom of the container allows to control the position of the air layers and potentially implement any desired shape at the interface. The oscillating sinking bubbles are used here as mechanical actuators to provide differentiate forcing in space. This simple setup provides an efficient way toward interface shape-control.

## 4. Transport phenomena on bended interfaces

It was shown recently that objects could float under liquid layer or at the surface of tilted interfaces [19, 26]. In the same manner, we now place plastic floaters of 1 cm at the bended surface. When initially placed far away from the air layer, the floater gets attracted toward the highest point of the interface above the air layer as shown in figure 4a and stops near the wall. The process is slow compared to the vibration, the floater horizontal velocity being typically

0.5 cm/s. During the motion, the immersed volume of the floater was observed to be constant up to measurement uncertainties (typically 5-10 % of the mean value), suggesting that Archimedes equilibrium is realized at each time. In the direction perpendicular to the images, the floater tends to systematically drift toward one of the wall of the container before starting its ascension, even when it is initially placed in the center.

In order to characterize this effect, we performed several experimental realizations for different excitation velocities and two different air volumes of Q = 10 ml and Q = 20 ml. During the ascent, the floater velocity $V_F$ is approximatively uniform as its horizontal coordinate varies linearly with time (see inset in figure 4b). The velocity is measured by performing a linear fit of the position along time. The results in figure 4b show that the floater velocity $V_F$ increases with the excitation velocity and the air layer volume. Depending on the parameters, floater velocity goes from 2 mm/s to more than 1 cm/s.

The origin of the slow motion of the floater can be found by looking at the small air bubbles motion close from the surface. We observe a slow secondary flow directed toward the air layer and the apparition of a vortex near the wall. This is reminiscent of vibration-induced vortices that have been observed in other experiments [27–29]. Those vortices are related to non-linear phenomenon occurring in the fluid. In our case, the surface flow associated to this vortex is responsible for the floater motion. The floater gets eventually trapped at the highest point of the interface while the small surrounding bubbles sink toward the vortex and are re-injected in the bulk.

Depending on the exact experimental configuration, the floater can be attracted or repelled as shown in figure 4c. When the air layer is trapped in the center of the container, two counter-rotating vertices are created. The induced flows tend to push the floater away from the highest point down to lowest point of the interface. As before, the floater follows the surrounding flow that is directed toward the lower point is this configuration.

The generation process of these vortices in our configuration is still to be understood. From a practical point of view, it nevertheless offers an original way to generate transport at a liquid surface.

## Conclusion

High frequency vibrations provide an efficient way to modify and manipulate liquid interface shape. Large air layers could be stabilized below the liquid through those vibrations. Those layers are acting as local vibration amplifiers, resulting in spatially inhomogeneous forcing and non-horizontal liquid interfaces. Tuning the bottom profile allows to control the position of the air layers in order to design the interface shape. Moreover, we have shown that secondary flows linked to vibration-induced vortices tends to move floaters placed at the interfaces toward preferred location. These results thus open the possibility of engineering the shape of liquid interfaces and control the transport of floating objects in specific regions.